\documentclass[conference,letterpaper]{IEEEtran}

\usepackage[status=final]{fixme}
\fxsetup{theme=color,author=initials,mode=multiuser,layout=inline}
\FXRegisterAuthor{mc}{amc}{Maciej}
\FXRegisterAuthor{jv}{ajv}{Javier}
\FXRegisterAuthor{og}{og}{Onur}

\usepackage[utf8]{inputenc}
\usepackage[T1]{fontenc}

\usepackage{booktabs}
\usepackage{graphicx}
\usepackage{xcolor}
\usepackage{amsmath,amsfonts,amssymb,amsthm}
\usepackage{verbatim}
\usepackage{subfig}
\usepackage{listings}
\usepackage{algorithmicx}
\usepackage[ruled,vlined]{algorithm2e}
\usepackage{algpseudocode}
\usepackage{tikz}
\usetikzlibrary{shapes,arrows,positioning}
\usepackage{svg}
\svgsetup{inkscapelatex=false, inkscapepath=svg-inkscape}
\usepackage{hyperref}
\hypersetup{
    bookmarks=false,        
    colorlinks=true,       
    pdfborder={0 0 0},     
}
\usepackage{url}
\usepackage{orcidlink}

\renewcommand{\url}[1]{\texttt{#1}}  

\usepackage[nameinlink]{cleveref}

\newtheorem{theorem}{Theorem}
\newtheorem{proposition}[theorem]{Proposition}

\newtheorem{corollary}{Corollary}

\theoremstyle{definition}

\theoremstyle{remark}

\newcommand{\Prob}[1]{\mathbf{P}\{#1\}}
\newcommand{\Exp}[1]{\mathbf{E}[#1]}
\newcommand{\bias}{\mathrm{bias}}

\DeclareMathOperator{\rank}{rank}

\title{Exact Bias of Linear TRNG Correctors: \\ Spectral Approach}

\author{
  \IEEEauthorblockN{Maciej Sk\'orski\orcidlink{0000-0003-2997-7539}\textsuperscript{1,2}, Francisco-Javier Soto\textsuperscript{3}\orcidlink{0000-0002-7217-7193}, and Onur G\"unl\"u\orcidlink{0000-0002-0313-7788}\textsuperscript{4, 5}}
    \IEEEauthorblockA{\textsuperscript{1}%
        University of Luxembourg, Luxembourg,
        maciej.skorski@gmail.com
    }
    \IEEEauthorblockA{\textsuperscript{2}%
        Czech Technical University in Prague, Czech Republic, 
    }
    \IEEEauthorblockA{\textsuperscript{3}%
        Rey Juan Carlos University, Móstoles, Spain, franciscojavier.soto@urjc.es
    }
    \IEEEauthorblockA{\textsuperscript{4}%
        Lehrstuhl für Nachrichtentechnik, Technische Universität Dortmund, Germany,
        onur.guenlue@tu-dortmund.de
    }
    \IEEEauthorblockA{\textsuperscript{5}%
        Information Theory and Security Laboratory (ITSL), Link{\"o}ping University, Sweden
    }
}

\usepackage[backend=biber,style=ieee]{biblatex}
\AtEveryBibitem{
  \clearlist{location}
  \clearlist{publisher}
  \clearfield{url}
  \clearfield{urlyear} 
}
\AtEveryCitekey{\clearlist{location}}
\begin{document}

\maketitle


\begin{abstract}

Using Fourier analysis, this paper establishes near-optimal security bounds for linear correctors commonly used in True Random Number Generators (TRNGs), expressed through code weight enumerators and input bias parameters. We provide the first near-tight bias characterization in total variation, by interpolating between optimal $\ell_\infty$ and $\ell_2$ norm results. Our bounds improve security assessments by an order of magnitude over previously known (overly conservative) estimates.

Across $\sim $20,000 codes, we examine fundamental trade-offs between compression efficiency, cryptographic security, and hardware complexity. Achieving 80-bit security with 10\% input bias typically requires sacrificing more than 50\% of the code rate and incurs increased hardware cost. This quantifies the inherent cost of randomness extraction in hardware TRNG implementations.
\end{abstract}

\section{Introduction}


True Random Number Generators (TRNGs) extract randomness from physical
phenomena, but their raw outputs typically contain statistical imperfections
(e.g., bias or correlations) that require post-processing to satisfy
cryptographic requirements.
Randomness extractors offer a principled way to correct such imperfections and have been studied extensively, from von Neumann's classical
procedure~\cite{von1963various} to later refinements by
Elias~\cite{eliasEfficientConstructionUnbiased1972},
Blum~\cite{blumIndependentUnbiasedCoin1986a},
and Zuckerman~\cite{zuckermanGeneralWeakRandom1990}.
However, hardware implementations favor simpler constructions, and physical
noise sources often satisfy stronger independence assumptions than those
required by general extractor theory. 
\mcnote{not noise sources, but constructions on top of them}
Linear correctors, introduced by Dichtl~\cite{dichtlBadGoodWays2007b}, strike an effective balance: they require only XOR gates and operate as
$Y = GX$,
where $X\in\mathbb{F}_2^n$ is the raw output, $Y\in\mathbb{F}_2^k$ is the
corrected output, and $G$ is a binary matrix. Their behaviour reduces to well-understood properties of the linear code generated by $G$, hence their popularity in practice~\cite{dichtlBadGoodWays2007b,lacharmePostProcessingFunctionsBiased2008b,lacharme_analysis_2009,hongchaozhouLinearExtractorsExtracting2011,grujicOptimizingLinearCorrectors2024,tomasiCodeGeneratorMatrices2017b}.

Most previous analyses of such correctors rely on $\ell_\infty$ bounds or on
summing individual input biases, techniques that often give security estimates far from the true total variation distance. In this work, we revisit the problem through a Fourier viewpoint, which expresses the distribution of $Y$ in a form determined by the weight enumerator of the code defined by $G$. This leads to exact formulas for the $\ell_\infty$ and $\ell_2$ distances to uniform and, by interpolation, significantly sharper estimates in total variation distance.

\subsection{Main Contributions}

This paper establishes near-optimal bounds for security of linear TRNG correctors, 
under the commonly used biased coin model\footnote{This model is in line with prior work~\cite{dichtlBadGoodWays2007b,lacharmePostProcessingFunctionsBiased2008b,lacharme_analysis_2009,hongchaozhouLinearExtractorsExtracting2011,grujicOptimizingLinearCorrectors2024,tomasiCodeGeneratorMatrices2017b} and is  reasonable for sources, such as ring oscillators and phase-locked loops.}, contributing:

\begin{itemize}

\item \textit{Fourier-analytic characterization.} Our Fourier methods yield optimal distance-to-uniformity formulas under $\ell_\infty$ and $\ell_2$ norms, expressed compactly through code weight enumerators.

\item \textit{Nearly tight $\ell_1$ bounds via $\ell_2$ interpolation.} Our bounds of the form
\begin{align}
\frac{W_G(\delta^2)-1}{W_G(\delta)-1} \leq \|\mathbf{P}_Y - \mathbf{P}_{U_k}\|_1 \leq \sqrt{W_G(\delta^2)-1}
\end{align}
improve over prior $\ell_{\infty}$-based estimates by orders of magnitude for practically interesting security levels.

\item \textit{Stable computation and empirical evaluation.}  
We illustrate numerically stable evaluations of $W_G(\delta)$ and $W_G(\delta^2)$ and compute the results on several families of linear correctors, highlighting the rate--security tradeoffs induced by different choices of $G$.\footnote{The source code is available via \href{}{https://osf.io/236yz/}}
\end{itemize}

\section{Related Work}

Several works have analyzed linear extractors and correctors. Lacharme derived an $\ell_\infty$ bound~\cite{lacharmePostProcessingFunctionsBiased2008b}, and
later gave a polynomial form under independent inputs~\cite{lacharme_analysis_2009}. Zhou et al.\ obtained an $\ell_1$
estimate~\cite{hongchaozhouLinearExtractorsExtracting2011}, and a
similar formulation appeared in Tomasi et
al.~\cite{tomasiCodeGeneratorMatrices2017b}. 
Recently, Gruji\'c \cite{grujicOptimizingLinearCorrectors2024} established a tight output min-entropy (equivalently, $\ell_\infty$) bound for linear correctors, in a formulation that already covers independent not necessarily identically distributed inputs, and studied trade-offs between security and implementation efficiency. 
In contrast, our main new
contributions are an exact $\ell_2$ characterization, nearly tight $\ell_1$ bounds
via interpolation, and a large-scale empirical study of rate--security--cost relations.

When these bounds are rewritten via the weight enumerator function and specialized to
biased independent coins, the resulting total variation estimates typically
decrease with the square root of the minimum-distance behavior of the code.
Building on this line of analysis, our work shows that the dependence can, in fact,
be made proportional to the full minimum-distance term, leading to substantially
sharper estimates in the small-bias regime while remaining fully expressible as a
function of the weight enumerator.
\mcnote{add more quantitative comparison}

\section{Preliminaries}


Define the bias of a binary random variable $Z$ as $\text{bias}(Z) = \mathbf{E}[(-1)^Z] \!=\! \mathbf{P}\{Z \!=\! 0\} \!-\! \mathbf{P}\{Z \!=\! 1\}$, and the XOR operation as $\oplus$. Linear correctors defined by the matrix $G \in \mathbb{F}_2^{k \times n}$ operate as $Y = G  X$, where $X = (X_i) \in \mathbb{F}_2^n$ and $Y = (Y_i) \in \mathbb{F}_2^k$ are $n$-bit input and $k$-bit output vectors. 
A linear code is a subspace of $\mathbb{F}_2^n$ and is defined by parameters $[n,k,d]$, where $n$ is the blocklength, $k$ is the dimension, and $d$ is the minimum distance (the smallest nonzero Hamming weight).
For a generator $G \in \mathbb{F}_2^{k \times n}$, the code is $C = \text{rowspan}(G)$ and codewords are represented as $c_S = \sum_{i \in S} G_i \in \mathbb{F}_2^n$, where $G_i$ are rows of $G$ and  $S \subseteq [k]$. The weight distribution counts codewords by Hamming weight $A_w = |\{c \in C : \|c\|_1 = w\}|$, giving the weight enumerator polynomial $W_G(x) = \sum_{w=0}^n A_w x^w$. When $k > n$ (overcomplete generators), $C \subseteq \mathbb{F}_2^n$ remains well-defined.
Weight distributions are available in repositories like OEIS~\cite{sloaneOnLineEncyclopediaInteger2007}, can be computed using Sage~\cite{developersSagemathSage952022} or Magma~\cite{bosmaMagmaAlgebraSystem1997}, or approximated for BCH and related codes using classical and probabilistic methods~\cite{tomlinsonBoundsErrorCorrectionCoding2017,kasamiApproximationWeightDistribution1985,krasikovSpectraBCHCodes1995,jainEstimatingWeightEnumerators2024}.

For any $S \subseteq [n]$, we define the \emph{parity function} $\chi_S(x) = (-1)^{\sum_{i \in S} x_i}$, which evaluates the parity of the bits indexed by $S$. These parity functions form an orthonormal basis for functions on boolean cube, allowing every function $f : \mathbb{F}_2^n \to \mathbb{R}$ to be expressed via the \emph{Fourier expansion}~\cite{odonnellAnalysisBooleanFunctions2014} $f(x) = \sum_{S \subseteq [n]} \hat{f}(S)\chi_S(x)$,
where the \emph{Fourier coefficients} are given by $\hat{f}(S) = 2^{-n}\sum_x f(x)\chi_S(x)$.
A key tool is \emph{Plancherel's theorem}, which shows that the $\ell_2$ norms in the time and frequency domains:
$2^{-n}\sum_x f(x)^2 = \sum_{S \subseteq [n]} \hat{f}(S)^2$.

Applied to probability mass functions, this yields the following useful characterization of $\ell_2$ distance

\begin{proposition}\label{prop:distance_isometry}
For any $Y$ over $k$ bits and uniform $U_k$, 
\begin{equation}
\label{eq}
\|\mathbf{P}_Y-\mathbf{P}_{U_k}\|_2^2  = 2^{k} \sum_{S\neq\emptyset}\widehat{\mathbf{P}_Y}(S)^2.
\end{equation}
\end{proposition}

\begin{proof}
Expanding the square $\ell_2$ norm, we obtain
\begin{align}
&\|\mathbf{P}_Y-\mathbf{P}_{U_k}\|_2^2 
= \sum_y \left(\mathbf{P}\{Y=y\}-2^{-k}\right)^2 \nonumber\\ 
&= \sum_y \mathbf{P}\{Y=y\}^2 - 2 \sum_y 2^{-k}\mathbf{P}\{Y=y\} + \sum_y 2^{-2k}  \nonumber\\ 
&= \sum_y \mathbf{P}\{Y=y\}^2 - 2^{-k}.
\end{align}
By applying Plancherel's theorem to $f(y)=\mathbf{P}\{Y=y\}$, 
$$
\sum_y \mathbf{P}\{Y=y\}^2 = 2^k \sum_{S \subseteq [k]} \widehat{\mathbf{P}_Y}(S)^2,
$$
and because $\widehat{\mathbf{P}_Y}(\emptyset)=2^{-k}$, we obtain
\begin{align}
\|\mathbf{P}_Y-\mathbf{P}_{U_k}\|_2^2  = 2^k \sum_{S\neq\varnothing} \widehat{\mathbf{P}_Y}(S)^2.
\end{align}
\end{proof}

\section{Main Results}

\subsection{Characterization of Output Distribution}

We begin with a general result that characterizes the outputs of linear correctors regardless of the input distribution. Our result extends beyond our model to other frameworks (Markov, hidden-Markov models) and holds for general matrices, including singular (i.e., with rank deficiency) matrices.

\begin{theorem}\label{thm:output_dist}
The probability of any output $y = Gx$ of the distribution $Y=G  X$ is equal to
\begin{align}
\Prob{Y=y} = 2^{-k}\sum_{S\subseteq[k]} \Exp{(-1)^{c_S\cdot X}}\; (-1)^{c_S\cdot x}
\end{align}
which can be equivalently written in terms of bias as
\begin{align}
\Prob{Y=y} = 2^{-k}\sum_{S\subseteq[k]} \bias ( c_S \cdot (X\oplus x) ).
\end{align}
\end{theorem}

\begin{proof}
The Fourier expansion of $f(y)=\Prob{Y=y}$ is
\begin{align}
&\Prob{Y=y}
= 2^{-k}\sum_{S\subseteq[k]} \widehat{f}(S)\,\chi_S(y)\nonumber
\\&= 2^{-k}\sum_{S\subseteq[k]} \Big(\sum_{y} \Prob{Y=y}\,\chi_S(y)\Big)\,(-1)^{\sum_{i\in S}y_i}.
\end{align}
By definition of expectation, we have
\begin{align}
\sum_{y} \Prob{Y=y}\,\chi_S(y)
= \Exp{\chi_S(Y)}
= \Exp{(-1)^{\sum_{i\in S}Y_i}}.
\end{align}
Since $Y_i=G_i X$, we have $(-1)^{\sum_{i\in S}Y_i} = (-1)^{c_S\cdot X}$.
Substituting these expressions into the Fourier expansion formula yields
\begin{align}
\Prob{Y=y}
= 2^{-k}\sum_{S\subseteq[k]} \Exp{(-1)^{c_S\cdot X}}\; (-1)^{c_S\cdot x}.
\end{align}
Using the definition of bias and the identity $c_S\cdot (X\oplus x) = c_S\cdot X + c_S\cdot x \bmod 2$, we have
\begin{align}
&\Prob{Y=y}
= 2^{-k}\sum_{S\subseteq[k]} \Exp{(-1)^{c_S \cdot (X\oplus x)}} \nonumber\\
&= 2^{-k}\sum_{S\subseteq[k]} \bias ( c_S \cdot (X\oplus x) ).
\end{align}
\end{proof}

Under the biased coin model, the output probabilities can be expressed as polynomials in input biases, yielding a particularly convenient computational form given below.

\begin{corollary}\label{thm:dist_iid}
Suppose $X_i \sim \mathrm{Bern}(p_i)$ are independent, and define $\delta_i = 1-2p_i = \bias(X_i)$.
Then, for any output $y = Gx$ of the random variable $Y = GX$, we have
\begin{align}
\Prob{Y=y} = 2^{-k} \sum_{S\subseteq[k]} \prod_{i}\big((-1)^{x_i}\delta_i\big)^{(c_S)_i}.
\end{align}
\end{corollary}

\subsection{Randomness Condensing - Discrepancy Under $\ell_{\infty}$ Norm}

Using ~\Cref{thm:dist_iid}, we next derive the exact characterization of the $\ell_{\infty}$ norm.
In cryptographic analysis, this characterizes effectiveness of the corrector as a min-entropy condenser, establishing security under unpredictability applications (e.g., digital signatures, message authentication codes).

\begin{theorem}\label{thm:linfty_bound}
Suppose $X_i\sim\mathrm{Bern}(p_i)$ are independent, and denote $\delta_i= 1-2p_i=\bias(X_i)$.
For $Y=GX$, where $G$ is $k \times n$, we have
\begin{align}
\|\mathbf{P}_{Y}\|_{\infty}
= 2^{-\mathrm{rank}(G)} \sum_{c \in \mathrm{rowspan}(G)} \prod_i |\delta_i|^{\,c_i},
\end{align}
and the maximum of $\Prob{Y=y}$ is achieved for $y=Gx$, where $x_i = \frac{1 - \mathrm{sign}(\delta_i)}{2}$ when $\delta_i \neq 0$ (and $x_i$ arbitrary when $\delta_i = 0$).
\end{theorem}

\begin{proof}
From ~\Cref{thm:dist_iid}, we have
$$
\Prob{Y=y} = 2^{-k} \sum_{S\subseteq[k]} \prod_{i}\big((-1)^{x_i}\delta_i\big)^{(c_S)_i}.
$$
To maximize over all $y = Gx$, we choose $x_i = \frac{1 - \text{sign}(\delta_i)}{2}$ when $\delta_i \neq 0$ (and $x_i$ arbitrary when $\delta_i = 0$).
This makes $(-1)^{x_i}\delta_i = |\delta_i|$ for all $i$, giving
$$
\max_y \Prob{Y=y} = 2^{-k} \sum_{S\subseteq[k]} \prod_i |\delta_i|^{(c_S)_i}.
$$
Since two subsets $S_1, S_2$ yield the same vector $c_{S_1} = c_{S_2}$ iff $S_1 \oplus S_2 \in \ker(G^T)$, each $c \in \mathrm{rowspan}(G)$ corresponds to exactly $2^{k - \rank(G)}$ subsets.
Therefore, we have
\begin{align}
&\max_y \Prob{Y=y}
= 2^{-k} \cdot 2^{k - \mathrm{rank}(G)} \sum_{c \in \mathrm{rowspan}(G)} \prod_i |\delta_i|^{c_i}  \nonumber\\
&= 2^{-\mathrm{rank}(G)} \sum_{c \in \mathrm{rowspan}(G)} \prod_i |\delta_i|^{c_i}.
\end{align}
\end{proof}

When input biases are jointly bounded, the maximum is achieved under i.i.d. distribution, yielding a compact formula in terms of the weight enumerator polynomial given below.

\begin{corollary}\label{cor:linfty_iid}
Among all independent coins $X_i\sim\mathrm{Bern}(p_i)$ with $|\bias(X_i)| \leq \delta$, the maximum $\ell_\infty$ norm of the corrector output is
\begin{align}
\max_{|\bias(X_i)| \leq \delta} \|\mathbf{P}_{Y}\|_{\infty} = 2^{-\mathrm{rank}(G)} W_G(\delta),
\end{align}
where $W_G(x) = \sum_{w=0}^{n} A_w x^w$ is the weight enumerator polynomial of $\mathrm{rowspan}(G)$,
The maximum is achieved for i.i.d. coins with $|\bias(X_i)| = \delta$.
\end{corollary}

\begin{proof}
From \Cref{thm:linfty_bound}, we have
\begin{align}
\|\mathbf{P}_{Y}\|_{\infty} = 2^{-\mathrm{rank}(G)} \sum_{c \in \mathrm{rowspan}(G)} \prod_i |\delta_i|^{c_i}.
\end{align}
To maximize over $|\delta_i| \leq \delta$, we need to maximize each product $\prod_i |\delta_i|^{c_i}$ subject to the constraints.
For any fixed $c$, this product is maximized when $|\delta_i| = \delta$ for all $i$ with $c_i = 1$, giving $\prod_i |\delta_i|^{c_i} = \delta^{\|c\|_1}$ where $\|c\|_1$ is the Hamming weight.

Therefore, denoting by $A_w$ the number of codewords of weight $w$ in $\mathrm{rowspan}(G)$, we obtain
\begin{align}
&\max_{|\delta_i| \leq \delta} \|\mathbf{P}_{Y}\|_{\infty}
= 2^{-\mathrm{rank}(G)} \sum_{c \in \mathrm{rowspan}(G)} \delta^{\|c\|_1} \nonumber\\
&= 2^{-\mathrm{rank}(G)} \sum_{w=0}^{n} A_w \delta^w = 2^{-\mathrm{rank}(G)} W_G(\delta).
\end{align}
\end{proof}

\subsection{Randomness Extraction - Discrepancy Under $\ell_2$ Norm}
Only full-rank matrices can be extractors: when $\rank(G)<k$, the output $Y=GX$ takes values only in the proper linear subspace $\mathrm{im}(G)\subsetneq \mathbb{F}_2^k$, and therefore is far from uniform on $\mathbb{F}_2^k$. Thus, throughout this subsection we assume $\rank(G)=k$.

Using ~\Cref{thm:dist_iid}, we next obtain the exact characterization under the $\ell_2$ norm.
For cryptography, this characterizes the performance of correctors as Rényi entropy extractors, which is known to imply security under indistinguishability-type applications such as encryption.

\begin{theorem}\label{thm:l2_distance}
Suppose $\rank(G)=k$. For $Y = GX$, where $X_i \sim \mathrm{Bern}(p_i)$ are independent with bias $\delta_i = 1-2p_i$, we have
\begin{equation}
\|\mathbf{P}_Y - \mathbf{P}_{U_k}\|_2^2
= 2^{-k}\!\sum_{c \in \mathrm{rowspan}(G)\setminus\{0\}} \prod_i (\delta_i^2)^{c_i}.
\end{equation}
\end{theorem}


\begin{proof}
By Proposition~\ref{prop:distance_isometry} and the Fourier formula from ~\Cref{thm:dist_iid}, we have
\begin{align}
&\|\mathbf{P}_Y-\mathbf{P}_{U_k}\|_2^2
= 2^{k} \sum_{S\not=\emptyset}\widehat{\mathbf{P}_Y}(S)^2 \nonumber\\
&= 2^{k} \sum_{S\not=\emptyset} \left(2^{-k} \prod_i \delta_i^{(c_S)_i}\right)^2 = 2^{-k} \sum_{S\not=\emptyset} \prod_i (\delta_i^2)^{(c_S)_i}.
\end{align}
Because $\rank(G)=k$, the map $S \mapsto c_S$ is a bijection between nonempty subsets of $[k]$ and nonzero codewords in $\mathrm{rowspan}(G)$. Hence, we have
\begin{align}
\|\mathbf{P}_Y-\mathbf{P}_{U_k}\|_2^2
= 2^{-k}\sum_{c \in \mathrm{rowspan}(G)\setminus\{0\}} \prod_i (\delta_i^2)^{c_i}.
\end{align}
\end{proof}

\begin{corollary}\label{cor:l2_iid}
Suppose $\rank(G)=k$. Among all independent coins $X_i\sim\mathrm{Bern}(p_i)$ with $|\bias(X_i)| \leq \delta$, the maximum $\ell_2$ distance to uniform distribution is
\begin{equation}
\max_{|\bias(X_i)| \leq \delta} \|\mathbf{P}_Y - \mathbf{P}_{U_k}\|_2
= \sqrt{2^{-k}(W_G(\delta^2)-1)}
\end{equation}
where $W_G(x) = \sum_{w=0}^n A_w x^w$ is the weight enumerator polynomial of $\mathrm{rowspan}(G)$.
The maximum is achieved by i.i.d. coins with $|\bias(X_i)| = \delta$.
\end{corollary}

\begin{proof}
From~\Cref{thm:l2_distance}, maximizing over $|\delta_i| \leq \delta$ gives $\prod_i (\delta_i^2)^{c_i} \leq \delta^{2\|c\|_1}$ with equality when all $|\delta_i| = \delta$.
Thus, we obtain
\begin{align}
&\max_{|\delta_i| \leq \delta} \|\mathbf{P}_Y-\mathbf{P}_{U_k}\|_2^2= 2^{-k}\sum_{c \in \mathrm{rowspan}(G)\setminus\{0\}} \delta^{2\|c\|_1}\nonumber\\
&= 2^{-k}(W_G(\delta^2) - 1).
\end{align}
Taking square roots completes the proof.
\end{proof}

\subsection{Randomness Extraction - Discrepancy Under $\ell_1$ Norm}

Typically in cryptography, the $\ell_2$ norm gives nearly sharp bounds on total variation (i.e., the $\ell_1$ norm). In our setting, these guarantees are information-theoretic and concern the output distribution itself, i.e., indistinguishability from the ideal uniform output $U_k$ without additional side information.

We first note that only full-rank matrices can be linear extractors. The reason is that rank deficiency leads to large Fourier coefficients which prevent proximity to uniformity.

\begin{proposition}[Linear extractors must be full-rank]\label{thm:l1_matrixrank}
If $G$ has rank deficiency, then for any input distribution $X$ we have $\|\mathbf{P}_Y - \mathbf{P}_{U_k}\|_{TV} \geq \frac{1}{2}$ and $\|\mathbf{P}_Y - \mathbf{P}_{U_k}\|_1 \geq 1$.
\end{proposition}

\begin{proof}
Suppose $G$ has rank deficiency, so there exists a non-empty subset $S \subseteq [k]$ such that $\sum_{i\in S} G_i = 0$.
Then, we have
\begin{align}
\chi_S(Y) = (-1)^{\sum_{i\in S} G_i \cdot X} = (-1)^0 = 1,
\end{align}
so $\Exp{\chi_S(Y)} = 1$.
For uniform $U_k$, we have $\Exp{\chi_S(U_k)} = 0$ since $S \neq \emptyset$. By the variational characterization of total variation, we obtain
$$
\|\mathbf{P}_Y - \mathbf{P}_{U_k}\|_{TV}
= \frac{1}{2}\sup_{f: \{0,1\}^k \to [-1,1]} \left|\Exp{f(Y)} - \Exp{f(U_k)}\right|.
$$
Taking $f = \chi_S$, gives
$$
\|\mathbf{P}_Y - \mathbf{P}_{U_k}\|_{TV}
\!\geq\! \frac{1}{2}\left|\Exp{\chi_S(Y)} - \Exp{\chi_S(U_k)}\right|
\!=\! \frac{1}{2}|1 - 0|
\!=\! \frac{1}{2}.
$$
Thus, $Y$ is far from uniformity, so $G$ cannot be a linear extractor.
\end{proof}

For full-rank matrices, we establish complementary bounds for extraction in terms of the $\ell_1$ norm (total variation distance). 

\begin{theorem}\label{thm:l1_bounds}
Suppose $G$ is full rank.
For independent $X_i \sim \mathrm{Bern}(p_i)$ with $|\bias(X_i)| \leq \delta$, we have
\begin{equation}
\|\mathbf{P}_{Y}-\mathbf{P}_{U_k}\|_1 \leq \sqrt{W_G(\delta^2) - 1}.
\end{equation}
For i.i.d. $X_i$ with $|\bias(X_i)| = \delta$, we have
\begin{equation}
\frac{W_G(\delta^2) - 1}{W_G(\delta) - 1} \leq \|\mathbf{P}_{Y}-\mathbf{P}_{U_k}\|_1 \leq \sqrt{W_G(\delta^2) - 1},
\end{equation}
where $W_G(x) = \sum_{w=0}^n A_w x^w$ is the weight enumerator polynomial of $\mathrm{rowspan}(G)$.
\end{theorem}

\begin{proof}
The upper bound follows from $\|x\|_1 \leq \sqrt{2^k}\|x\|_2$ and Corollary~\ref{cor:l2_iid}, which gives
\begin{align}
\|\mathbf{P}_{Y}-\mathbf{P}_{U_k}\|_2 \le \sqrt{2^{-k}(W_G(\delta^2)-1)}.
\end{align}
Hence, we have
\begin{align}
&\|\mathbf{P}_{Y}-\mathbf{P}_{U_k}\|_1
\leq \sqrt{2^k}\,\sqrt{2^{-k}(W_G(\delta^2)-1)}\nonumber\\
&= \sqrt{W_G(\delta^2)-1}.
\end{align}

For the lower bound, we use $\|x\|_1 \ge \|x\|_2^2/\|x\|_\infty$.
By Corollary~\ref{cor:l2_iid}, we have
$\|\mathbf{P}_Y-\mathbf{P}_{U_k}\|_2^2 = 2^{-k}\bigl(W_G(\delta^2)-1\bigr)$.
In addition, by~\Cref{thm:dist_iid,cor:linfty_iid},
$ \|\mathbf{P}_Y-\mathbf{P}_{U_k}\|_\infty
=2^{-k}\bigl(W_G(\delta)-1\bigr). $
Consequently, we have
$\|\mathbf{P}_Y-\mathbf{P}_{U_k}\|_1
\geq
\frac{2^{-k}\bigl(W_G(\delta^2)-1\bigr)}
     {2^{-k}\bigl(W_G(\delta)-1\bigr)}
=
\frac{W_G(\delta^2)-1}{W_G(\delta)-1}.$
\end{proof}

\subsection{Optimality Discussion}

For small input bias $\delta \ll 1$, the weight enumerator polynomial is dominated by the
minimum-weight codewords, so $W_G(\delta)-1 \approx A_d \delta^d$, where $d$ is
the minimum distance. Thus, we have
$\|\mathbf{P}_Y - \mathbf{P}_{U_k}\|_1 = \Theta(\delta^d)$,
i.e., the total variation distance decays as $\delta^d$, up to a constant
factor depending only on $A_d$. Furthermore, many "good" codes have approximately binomial weights~\cite{geilSecondWeightGeneralized2008,kasamiApproximationWeightDistribution1985,bargRandomCodesMinimum2002,debris-alazardCodebasedCryptographyLecture2023}, so that $A_j = O\left(2^{k-n}\binom{n}{j}\right)$ and 
 the entropy bound $\binom{n}{d} \leq 2^{nh(d/n)}$ and Gilbert--Varshamov's bound $k/n \leq 1 - h(d/n)$ give
 $A_d \leq  O(1)$, making our bounds tight up to universal constants. 
Specifically, for random linear codes, which nearly meet the Gilbert--Varshamov bound with $k/n \approx 1 - h(d/n)$~\cite{bargRandomCodesMinimum2002}, to achieve security level $\epsilon = 2^{-s}$ we need $\delta^d \leq \epsilon$, giving $d \geq s/\log_2(1/\delta)$. This establishes the rate-security tradeoff $\displaystyle k \approx  n - O\left(\frac{s}{\log_2(1/\delta)}\right)$ on entropy (compare with \Cref{fig:security-tradeoffs}).


\section{Numerical Evaluations}\label{sec:num-eval}

We first illustrate our bounds on four representative codes to compare them with previous bounds. For Reed--Muller (RM) codes $\mathrm{RM}(r,m)$ with length $n = 2^m$, dimension
$k = \sum_{i=0}^{r} \binom{m}{i}$, and minimum distance
$d = 2^{m-r}$~\cite{abbeReedMullerCodes2021}, we use
$\mathrm{RM}(3,8)$ with parameters $[256,93,32]$ and $\mathrm{RM}(3,7)$ with
$[128,64,16]$.
For BCH codes of length $n = 2^m - 1$ and designed distance $d$, we consider
$[127,50,27]$ and $[255,47,85]$.
Weight enumerator polynomials are taken from OEIS sequences A018895, A146953, A097479, and
A151933. For these codes, we compare our new $\ell_1$ bounds with previous estimates,
rewritten in terms of $W_G$. \Cref{fig:security-bounds-rm,fig:security-bounds-bch} show that, for a given
input bias $\delta$, our bounds yield strictly tighter security assessments,
often by a factor close to an order of magnitude in the range of practical
parameters. 
\Cref{fig:security-bands} depicts both the upper and lower bounds from
\Cref{thm:l1_bounds} and indicates that the gap is small for
``good'' codes, so $\sqrt{W_G(\delta^2)-1}$ is a good proxy for exact total
variation.

\begin{figure}[h!]
    \centering
    \includegraphics[width=0.95\columnwidth]{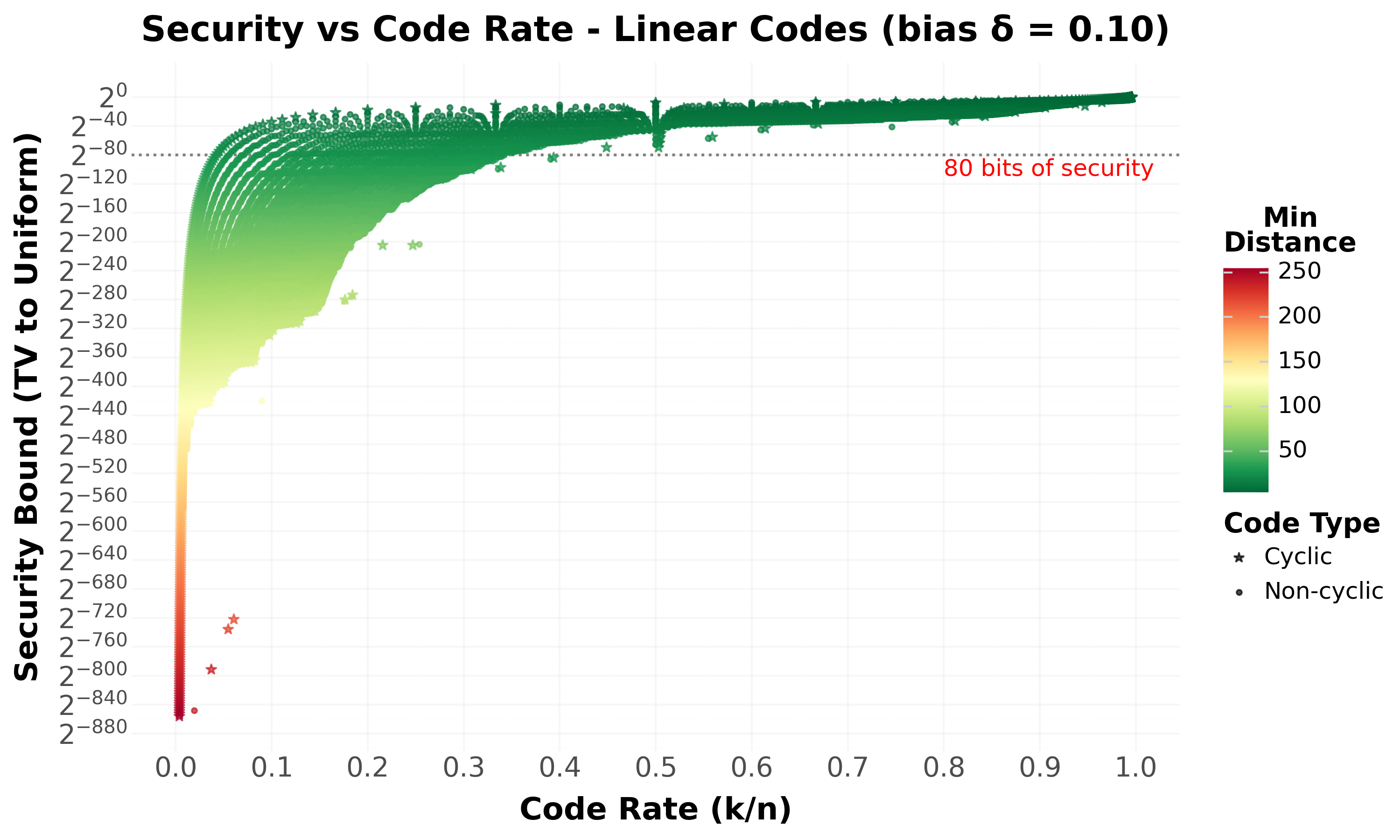}
    \caption{Rate-security tradeoff for codes from~\cite{grujicOptimizingLinearCorrectors2024}, showing rate $R=k/n$ vs. security $s$ obtained with our method.}
    \label{fig:security-tradeoffs}
\end{figure}

To study the rate--security tradeoff, we use a large dataset of roughly
$20{,}000$ linear codes  from~\cite{grujicOptimizingLinearCorrectors2024}.
For each code, and for fixed input bias $\delta$ and target security $s$, we
test whether
\begin{align}
  \sqrt{W_G(\delta^2)-1} \le 2^{-s}.
\end{align}
Evaluations of $W_G$ are carried out in the
log-domain with vectorized \texttt{log-sum-exp}, to avoid  overflow and underflow for small $\delta$ and large weights.
The resulting scatter plot in~\Cref{fig:security-tradeoffs} shows that maintaining $80$-bit security typically forces the rate down to roughly $0.3$--$0.5$, depending on the code family. Encouragingly, we observe quite often cyclic codes at the Pareto frontier, underscoring their favorable balance of hardware implementation cost, security, and rate.

\begin{figure}[h!]
    \centering
    \includesvg[width=0.95\columnwidth]{figures/security_bounds_rm_codes}
    \caption{Security bounds comparison for RM codes.}
    \label{fig:security-bounds-rm}
\end{figure}

\begin{figure}[h!]
    \centering
    \includesvg[width=0.95\columnwidth]{figures/security_bounds_bch_codes}
    \caption{Security bounds comparison for BCH codes.}
    \label{fig:security-bounds-bch}
\end{figure}

\begin{figure}[h!]
    \centering
    \includesvg[width=0.95\columnwidth]{figures/security_bands_rm}
    \caption{Tightness of our security bounds from \Cref{thm:l1_bounds}.}
    \label{fig:security-bands}
\end{figure}

\begin{figure}[h!]
\includegraphics[width=0.90\columnwidth]{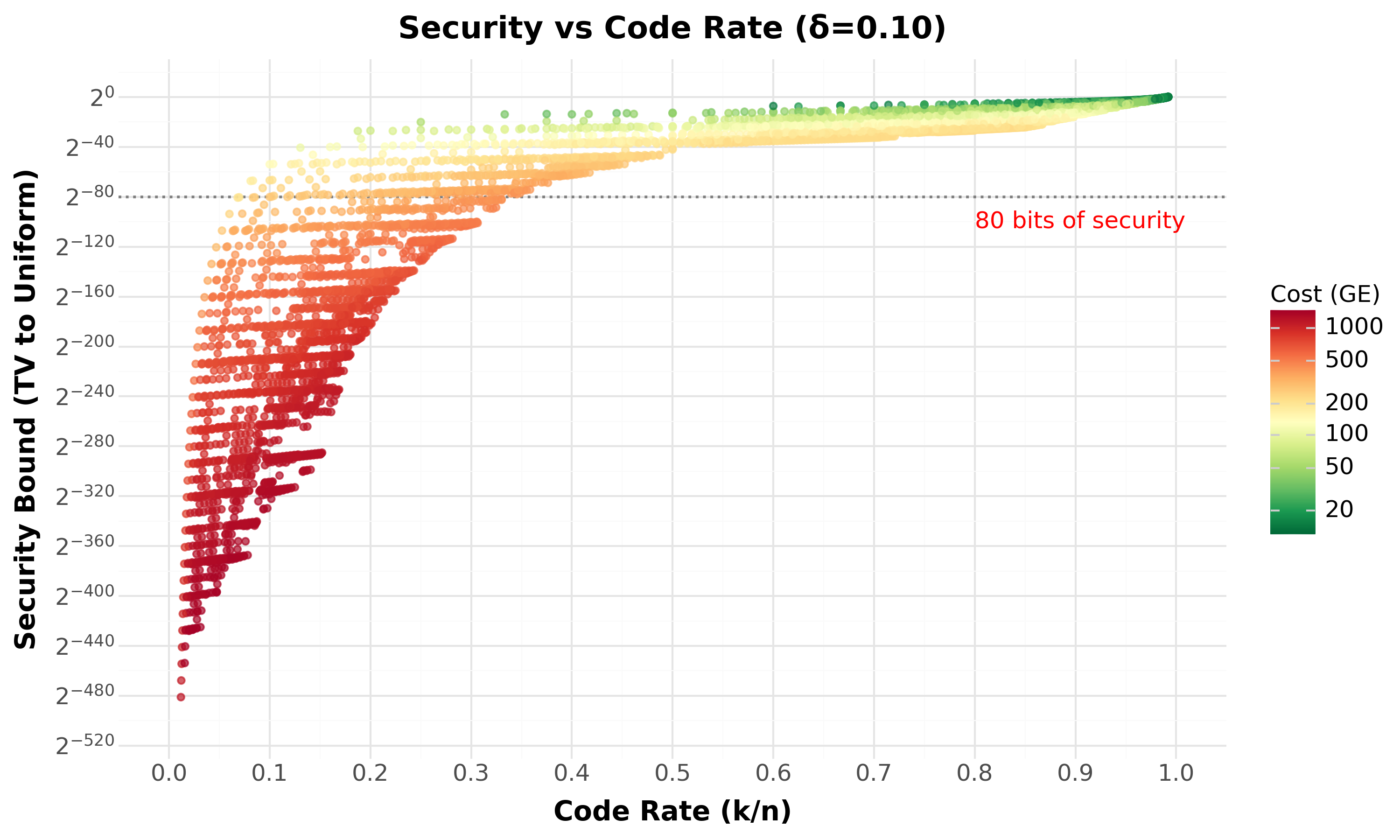}
\caption{Security vs. Rate vs. Implementation cost.}
\label{fig:cost}
\end{figure}

Finally, we estimate hardware cost. Computing $Y = Gx$ may require considerable chip area~\cite{dichtlBadGoodWays2007b}. We measure cost in Gate Equivalents (GE), using the standard values 2.67 GE per XOR gate and 6 GE per register~\cite{kwokComparisonPostProcessingTechniques2011b,poschmannLightweightCryptography2009}. These GE values serve only as normalized proxies for relative comparison, not as post-layout area estimates for a specific technology. For each code, we evaluate both the generator matrix $G$ and the parity-check matrix $H$ in systematic form, and keep the cheaper implementation. Results in~\Cref{fig:cost} show that higher security generally comes at higher hardware cost.

\section{Conclusion}

We derived near-optimal security bounds for linear TRNG correctors using Fourier analysis, establishing exact $\ell_2$ and $\ell_\infty$ formulas and nearly tight $\ell_1$ bounds via interpolation. All bounds are expressed through code weight enumerators, providing a unified coding-theoretic framework. Our bounds significantly sharpen previous results, particularly for small bias: total variation distance scales as $\delta^d$ (where $d$ is the minimum distance) rather than its square root as in prior $\ell_\infty$-based bounds. Experiments with RM, BCH, and other codes showed order-of-magnitude improvements while making rate–security tradeoffs explicit, e.g., achieving $80$-bit security at $\delta=0.1$ requires substantial rate loss.

In future, we will consider integrating hardware constraints (area, power) for practical use cases and extending our results to other important code families, such as polar codes~\cite{arikan2009channel}, e.g., by using exact or approximate weight enumerators~\cite{PolarWEF}.


\section*{Acknowledgment}
This work was supported by Czech project CROP CZ.02.01.01/00/22\_011/0008569, Poland's Excellence Initiative -- Research University (IDUB) program, the ``PREDOCT2022-006'' program, Swedish Foundation for Strategic Research (SSF), and BMFTR 6GEM+ Transfer Hub under Grants 16KIS2412 and 16KISS005.

\newpage

\printbibliography


\end{document}